\documentclass{article}
\usepackage{emulateapj,epsfig,onecolfloat}

\begin{document}
\twocolumn
[
\title{The DEEP2 Galaxy Redshift Survey: Spectral classification \\ 
of galaxies at $z\sim1$}

\author{ Darren S. Madgwick\altaffilmark{1}, 
Alison L. Coil\altaffilmark{1}, 
Christopher J. Conselice\altaffilmark{2}, 
Michael C. Cooper\altaffilmark{1}, 
Marc Davis\altaffilmark{1}, 
Richard S. Ellis\altaffilmark{2}, 
Sandra M. Faber\altaffilmark{3}, 
Douglas P. Finkbeiner\altaffilmark{4}, 
Brian Gerke\altaffilmark{1}, 
Puragra Guhathakurta\altaffilmark{3,5}, 
Nick Kaiser\altaffilmark{6},
David C. Koo\altaffilmark{3}, 
Jeffrey A. Newman\altaffilmark{1}, 
Andrew C. Phillips\altaffilmark{6}, 
Charles C. Steidel\altaffilmark{2}, 
Benjamin J. Weiner\altaffilmark{3},
Christopher N. A. Willmer\altaffilmark{3}, 
R. Yan\altaffilmark{1}
(the DEEP2 Survey Team)
}

\altaffiltext{1}{Department of Astronomy, University of California,
Berkeley, CA 94720}
\altaffiltext{2}{Department of Astronomy, California Institute of
Technology, Pasadena, CA 91125}
\altaffiltext{3}{University of California Observatories/Lick
Observatory, Department of Astronomy and Astrophysics, University of
California, Santa Cruz, CA 95064}
\altaffiltext{4}{Princeton University Observatory, Princeton, NJ 08544}
\altaffiltext{5}{Herzberg Institute of Astrophysics, National Research
Council of Canada, 5071 West Saanich Road, Victoria, B. C., Canada V9E 2E7}
\altaffiltext{6}{Institute for Astronomy, University of Hawaii,
Honolulu, HI 96822}

\begin{abstract}
We present a Principal Component Analysis (PCA)-based spectral
classification, $\eta$,  for the first 5600 galaxies
observed in the DEEP2 Redshift Survey.  This parameter
provides a very pronounced separation between
absorption and emission dominated galaxy spectra -- corresponding to
passively evolving and actively star-forming galaxies in the survey
respectively.  In addition it is shown that despite the high
resolution of the observed spectra, this parameter alone can be used
to quite accurately reconstruct any given galaxy spectrum, suggesting
there are not many `degrees of freedom' in the observed spectra of this
galaxy population.
It is argued that this form of classification, $\eta$, will be particularly 
valuable in
making future comparisons between high and low-redshift galaxy
surveys for which very large spectroscopic samples are now readily
available, particularly when used in conjunction with high-resolution spectral
synthesis models which will be made public in the near future.  
We also discuss the relative advantages of this
approach to distant galaxy classification compared to other
methods such as colors and morphologies.  Finally, we compare
the classification derived here with that adopted for the 2dF Galaxy
Redshift Survey and in so doing show that the two systems are very
similar.  This will be particularly useful in subsequent analyses when
making comparisons between results from each of these surveys to study
evolution in the galaxy populations and large-scale structure.
\end{abstract}

\keywords{Galaxies: high-redshift --- galaxies: evolution}
]

\section{Introduction}

The classification of galaxies is of fundamental importance for
understanding galaxy populations, and for this reason is a
very important aspect of any galaxy redshift survey.  
Having a data set of
many thousands of galaxy spectra allows one to test the
validity of galaxy 
formation and evolution scenarios with unprecedented accuracy.  However,
the sheer size of the full spectral data set presents its 
own unique problems.
In order to make such a galaxy data set more 
`digestible' some form of data
compression is necessary, whether this be through the adoption of
morphological segregation, colors or some other
compression/classification scheme.
If these quantities (and their associated distributions) can be
determined consistently over a wide 
range of redshifts, they can be compared with theoretical predictions and
simulations, and hence set constraints on scenarios for
galaxy evolution.  This will be especially true if consistent
classification regimes can be used for both the high redshift ($z\sim1$)
surveys currently underway (the  
DEEP2 Galaxy Redshift Survey,
Davis et al. 2002 and the VIRMOS-VLT Deep Survey, Le Fevre et
al. 1999) -- and the large $z\sim 0$ surveys
now approaching full 
completion; the Sloan Digital Sky Survey (SDSS, Strauss et al. 2002)
and the 2dF Galaxy Redshift Survey (2dFGRS, Colless et al. 2001)

A number of
different approaches to the
classification of galaxy spectra have been adopted for local galaxy surveys.
These include the calculation of
rest-frame colors (e.g. Strateva et al. 2001);
principal component analysis (PCA) based spectral classifications
(e.g. Connolly et al. 1995; Bromley et al. 1998; Folkes et al. 1999;
Madgwick et al. 2002; de Lapparent et al. 2003), and other more sophisticated discriminations
(e.g. Heavens, Jimenez \& Lahav 2000; Slonim et al. 2001), based upon
information theory.  The
underlying theme of all these alternative methods is that they
characterize the galaxy population exclusively in terms
of their observed spectra.  The work presented here is the first
attempt to apply one of these methods (PCA) to the classification of
such a distant sample of galaxies.

\subsection{The role of spectral classification}

There are three methods which have generally proved to be
popular for the classification of galaxies:
morphological segregation, rest-frame colors and direct spectrum based
classifications.  Each of these methods has its own unique drawbacks and
advantages.  

To understand galaxy evolution
out to redshifts of $z\ga1$, it is essential to have a consistent 
implementation of
these classifications over a wide range of look-back times.  
For this reason it can be argued that morphological
segregation -- although perhaps the most natural form of classification -- 
may not be the optimal solution to adopt over such a large range of
redshifts.  This is due
to both the degradation of morphology with redshift and the absence of
a robust and 
repeatable methodology to perform this classification (see e.g. Conselice
2003 for further discussion and possible solutions to this
situation).  For this reason we focus in this paper on alternative
classification methods to morphology, which we hope will complement
earlier studies based on this method, whilst at the same time
providing a new perspective by more directly reflecting the physical
properties of each galaxy.

The remaining two options -- rest-frame colors and spectral `types' --
are linked, in that they both provide
some compressed representation of the observed galaxy's spectral
energy distribution (SED), hence providing a relatively direct insight
into physical processes such as star-formation currently occurring in
each galaxy.  However, in terms of how each is
calculated for high-z galaxies there are significant differences
which are important to understand.

The main complication
for the calculation of rest-frame colors is
the accurate determination of $k$-corrections to account for the
different pass-bands sampled by each photometric filter at different
redshifts.  These generally cannot be estimated from the observed
spectrum, but rather must be determined by matching each observed
galaxy to a set of template SEDs 
with full rest-frame wavelength coverage.

In the case of spectral classification, one must consider that
the rest-frame wavelength coverage varies with the
redshift of the galaxy, hence to adopt a uniform
classification over a large range of redshifts one needs to
significantly restrict the rest-frame wavelength range considered
(e.g. by focusing in on particular line features
through equivalent width measurements) or to determine some way of 
filling in the gaps that are not observed in each spectrum.

These two problems are very closely related in that they express the need 
to determine the form of a galaxy's SED over a wavelength interval
that is not necessarily observed.  
In the case of $k$-corrections
only $\sim10$ template SEDs are available for this task (Kinney et
al. 1996; Coleman, Wu \& Weedman 1980); reddening is manually
incorporated and we have little control over the wavelength intervals
involved.
However as will be shown in this paper, with principal
component analysis (PCA), we can achieve this task for spectral
classification, using the observed spectra themselves as
templates, giving us $\sim N_{\rm gal}$ galaxy templates, with which
to interpolate or extrapolate a given observed spectrum.  In
addition, because we have much more control over the wavelength
interval adopted for the classification we can modify the analysis to
use only the parts of the observed spectra which have the most uniform
rest-frame wavelength  coverage.  

It is primarily for this latter reason that adopting PCA-based spectral
types will provide a
classification scheme that is particularly uniform over the large range of
redshifts encountered in the DEEP2 redshift survey, hence 
providing a robust probe of evolution in the galaxy
population.  
In addition we note that such methods of classification are timely, in
that there 
is now a considerable body of low redshift spectroscopic data from
e.g. 2dFGRS and SDSS, which enables us to make detailed comparisons.
We note that PCA-based classifications do suffer from one
particular drawback, which is that they are not as straightforward to
interpret as other classifications, and this is an issue we will
attempt to address in this paper.

\begin{figure*}
\begin{center}
\epsfig{file=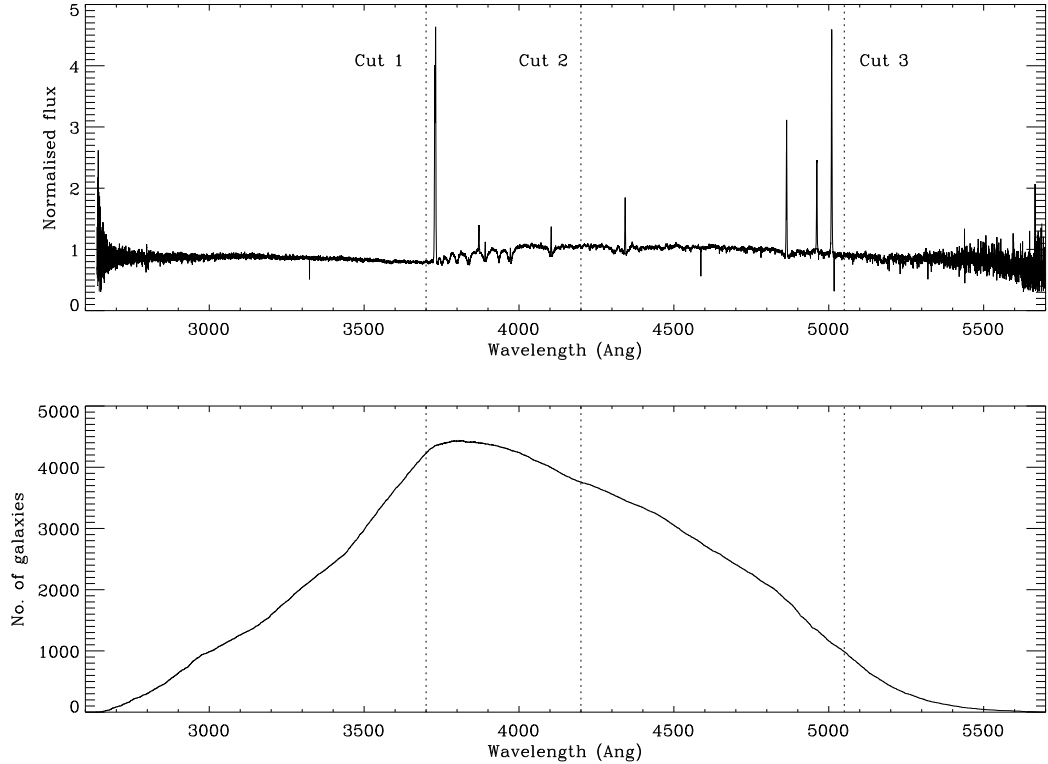,angle=90,width=5.5in}
\caption{The mean rest-frame spectrum for all galaxies ($z>0.6$)
observed to date 
in the DEEP2 Redshift Survey.  The top panel shows the mean spectrum
in units of normalized flux, while the bottom panel shows how many
galaxies have contributed to each wavelength channel (which is
determined by the redshift and wavelength coverage of each galaxy).
Also shown (dotted lines) are 
the wavelength ranges considered in our PCA analyses, one
spans Cut1 to Cut 2 ($3700-4200$\AA), the second Cut 1 to Cut 3
($3700-5050$\AA).\label{fig:aver}}
\end{center}
\end{figure*}

The outline of this paper is as follows:  In Section~2 we briefly discuss the
DEEP2 Redshift Survey data that we will use in this analysis.
Section~3 describes the implementation of PCA we have adopted for this
paper, and gives a discussion as to why this particular formalism has
been used.  In particular, issues regarding the restframe wavelength
coverage are discussed in some detail.  In Section~4 we discuss the
method of spectral classification we will adopt, based upon the
results of the PCA we have implemented.  This spectral classification
is contrasted with that of the 2dF Galaxy Redshift Survey in
Section~5, in which the selection effects of the two surveys are also
discussed.  We then conclude this paper in Section~6 with a brief
discussion as to future applications of this work.

\section{DEEP2 galaxy spectra}

In its first season of observations (August -- October, 2002), the
DEEP2 Redshift Survey has already accurately 
measured the redshifts of $\sim5600$ galaxies, out of a proposed total
of 60,000. 
These galaxies have been pre-selected to have $z\ga0.7$ and an 
$R_{\rm AB}$ limiting magnitude
of 24.1, from a set of $B$,$R$ and $I$ CFHT 12k x 8k 
images covering $\sim4$ deg$^2$ on the
sky.  Foreground galaxies have been excluded using a simple photometric
cut, based upon the observed $R-I$ and $B-R$ color of
each galaxy (Davis et al. 2002).  In this paper we make use of all
this data from the first observing season.

A sophisticated automated pipeline has been developed to efficiently extract 
and reduce the spectra
observed in the survey, details of which 
will be presented by Davis et al. (2003). 
The observed spectra themselves are taken at high resolution 
($R\sim5000$) using the DEIMOS spectrograph mounted on the 10-m Keck-II
telescope (Davis \& Faber, 1998), 
and generally span the wavelength range of $6400 <
\lambda < 9200$\AA.  For galaxies with $z>0.7$ this allows us to very
accurately measure the redshift of each galaxy, particularly when the
resolved [O{\sc ii}] doublet is present 
(out to a redshift of $z=1.5$).  Absorption based redshifts are also
readily determined, primarily using the Ca H+K features which are
visible to redshifts of $z\sim1.3$.

All spectra used in this analysis have been corrected for
the DEIMOS throughput efficiency of the 1200 line/mm grating used in
the survey\footnote{see {\tt
http://www.ucolick.org/$^\sim$ripisc/Go1200.html}}, and
are presented in
units of counts/pixel/hour.  Note that the flux calibration is only
approximate at present ($\sim$10\%), however we have found the components
derived from our PCA to be robust to the exact value of the throughput
efficiency. 
Additionally, the spectra have been normalized to have a mean count
of unity, after being minimally smoothed (3
pixels $\sim0.9$\AA\ observed-frame, $\sim0.5$\AA\ restframe) according to the
inverse of their estimated variance (in order to remove sky
residuals and other artifacts remaining from the spectral
reduction).

\begin{figure}
\plotone{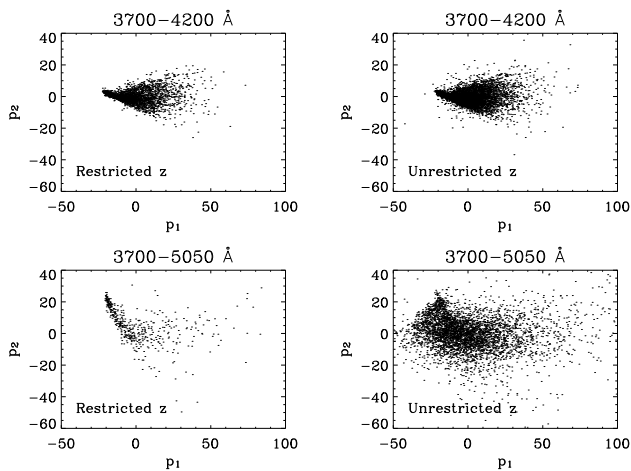}
\caption{The first two projections ($p_1$ and $p_2$) are shown for
both the PCA analyses.  The top two panels show the projections for
the PCA defined using only the 3700-4200 \AA wavelength range, while the
bottom cover 3700-5050
\AA.  In each case the left panel shows the projections for
only those galaxies which span the full rest-frame wavelength range
used, whereas the right panel shows the projections for all DEEP2
galaxies, regardless of their restframe wavelength coverage.
It can be seen that the PCA is consistent in both samples for the
shorter wavelength 
range adopted, but not for the larger range.  This is also found to be
true of higher principal components $p_3$, $p_4$ etc.\label{fig:pcs}}
\end{figure}

\section{Principal Component Analysis}

The spectral classification presented here is based upon a Principal
Component Analysis (PCA) of the DEEP2 galaxy spectra.  
PCA is a powerful technique, allowing us to easily visualize
and quantify
a multi-dimensional population in terms of just a handful of
significant components.  It does this by identifying the components of 
a data set (in this case the galaxy spectra) which are the most discriminatory
between each galaxy (where the
significance is determined in terms of its
contribution to the variance over the entire sample).
This allows us to identify just the most
significant components for future use. 
It is clear from such a formalism that any clustering in the
PCA space is indicative of distinct sub-populations within the
sample. 

PCA has
been used with considerable success by a variety of authors
to deal with large multi-dimensional data sets. 
Several similar mathematical formulations of PCA 
for galaxy spectra have appeared in the literature, in particular
Connolly et al. (1995); Bromley et al. (1998); Galaz \& de Lapparent (1998);
Folkes et al. (1999). The most significant difference between the
various techniques is that some methods utilize
mean-subtracted covariance matrices for the PCA (by subtracting the
mean of the normalized spectrum of each galaxy), while others do not.
This has little effect on the subsequent analysis
since the latter methods simply yield a 
mean spectrum as the first component.  

In this paper we present a new variation on previous formalisms 
for carrying out the PCA of galaxy spectra, designed to accommodate the
features of our data set that make
the standard method very difficult to apply.
In particular there are two complications that must be addressed
to analyze the DEEP2 galaxy
spectra: the first is the very high resolution of the observed
spectra which require extremely time-consuming matrix computations.  
The second is that the observed spectra
    cover very different rest-frame wavelength ranges once  
    de-redshifted, and also have a number of effective 'gaps' present
    due to the 
    presence of strong night-sky lines. For this reason a method of
    performing interpolations and (small) extrapolations is required
    to ensure the classification is uniform.

\subsection{Implementing PCA for high-resolution data}

A significant drawback to implementing PCA
on large or very high-dimensional data sets is the required
computation time, particularly to 
determine the covariance matrix.  For $n$ galaxies - each
with $p$ spectral channels - this requires $O(np^2)$ operations.
Given that each DEEP2 spectrum contains $O(10^4)$ channels this would
be very time consuming indeed if the full spectral
resolution is used.

Fortunately, it is possible to solve for the PCA eigenspectra without
calculating the entire covariance matrix.  The key is
developing a probabilistic formalism for the PCA,
compatible with an expectation-maximization (E-M) algorithm (see Roweis 1997
and Tipping \& Bishop 1999).  Adopting such a formalism allows one to
solve for the first $k$ eigenspectra in only $O(npk)$
operations.  It has been shown that this method for performing PCA is
robust, 
in that it has only one stationary point that is not a saddle point,
which guarantees there will be no false convergences (Tipping \&
Bishop 1999).   

Computationally, the E-M algorithm proceeds in iterations of two
steps.  The $k$ eigenvectors to be calculated
are assumed to be spanned by the columns of a $p\times k$
matrix $\bf{C}$.  We start by making an initial (random) guess for the
columns of this matrix and use this to determine the $k\times n$
matrix, ${\bf{X}}$, of $k$ `states' for each galaxy. These states
correspond to the principal component projections of each galaxy in
the non-orthogonal space defined by the columns of $\bf{C}$.
These states are then used in conjunction with the original $p\times
n$ data
matrix, $\bf{Y}$, to make a better estimate for $\bf{C}$.  This
proceeds until convergence.  These steps can be summarized as,
\begin{enumerate}
 \item Step I: $$ {\bf{X}} = ({\bf{C}}^T{\bf{C}})^{-1}{\bf{C}}^T{\bf{Y}} $$
 \item Step II: $${\bf{C}}^{\rm new} = {\bf{YX}}^T({\bf{XX}}^T)^{-1}$$
\end{enumerate}
Once the algorithm converges the columns of ${\bf{C}}$ will span the
space of the first $k$ eigenvectors.  This can therefore be used to
construct the orthogonal basis that defines the principal components
and their projections.  We note that this method for PCA gives
identical principal components to the other (simpler) formalisms
discussed previously.  The sole reason we adopt an E-M based
approach is to make the PCA efficient for the large and very high
resolution data set we are using.
Throughout the remainder of this paper we will denote 
the eigenvectors determined by the PCA (herein eigenspectra) as
$\bf{P_1}$,$\bf{P_2}$ etc. and the projections 
onto these axes by $p_1$, $p_2$ etc.

\begin{figure}
\begin{center}
\epsfig{file=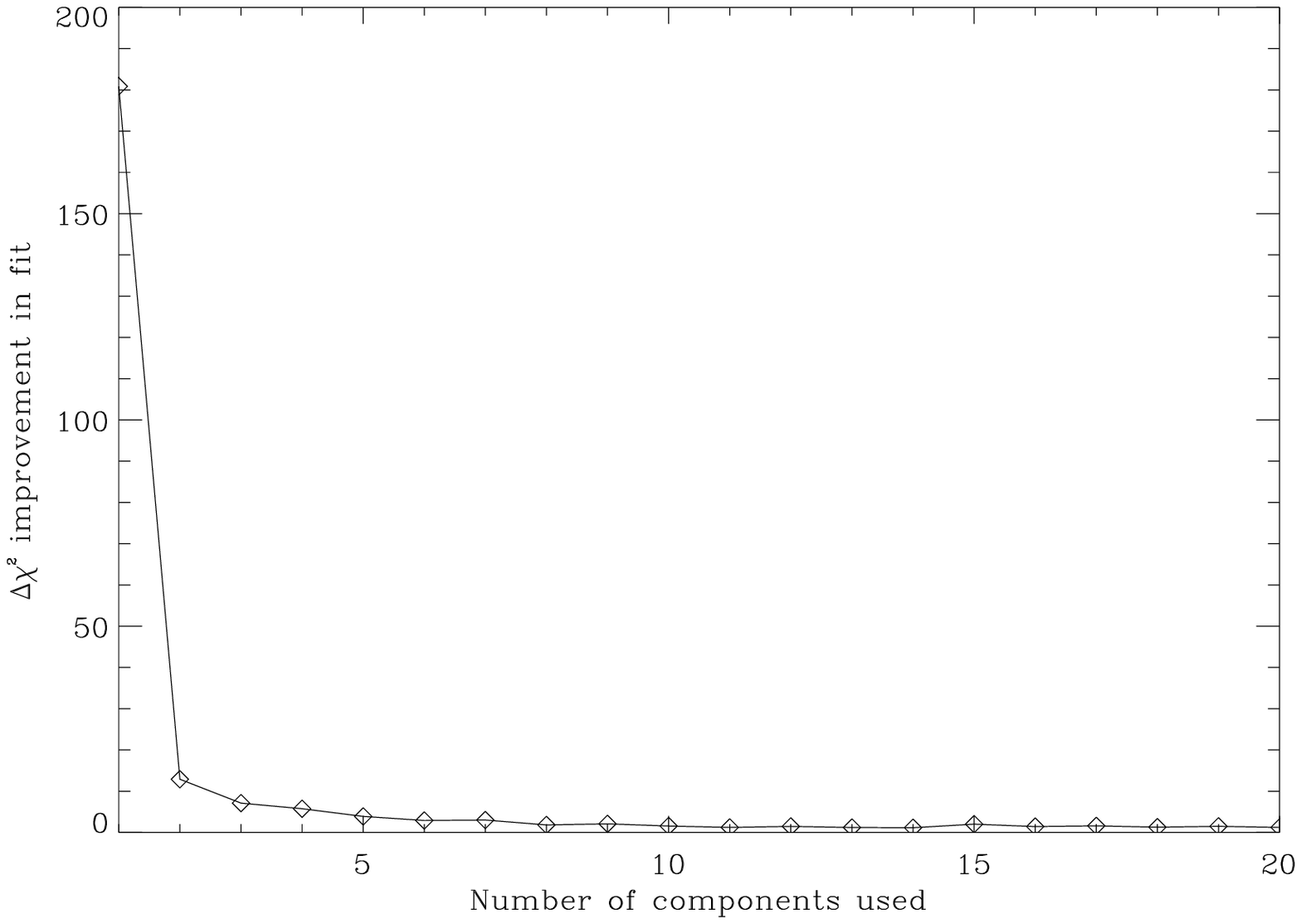,width=3in}
\epsfig{file=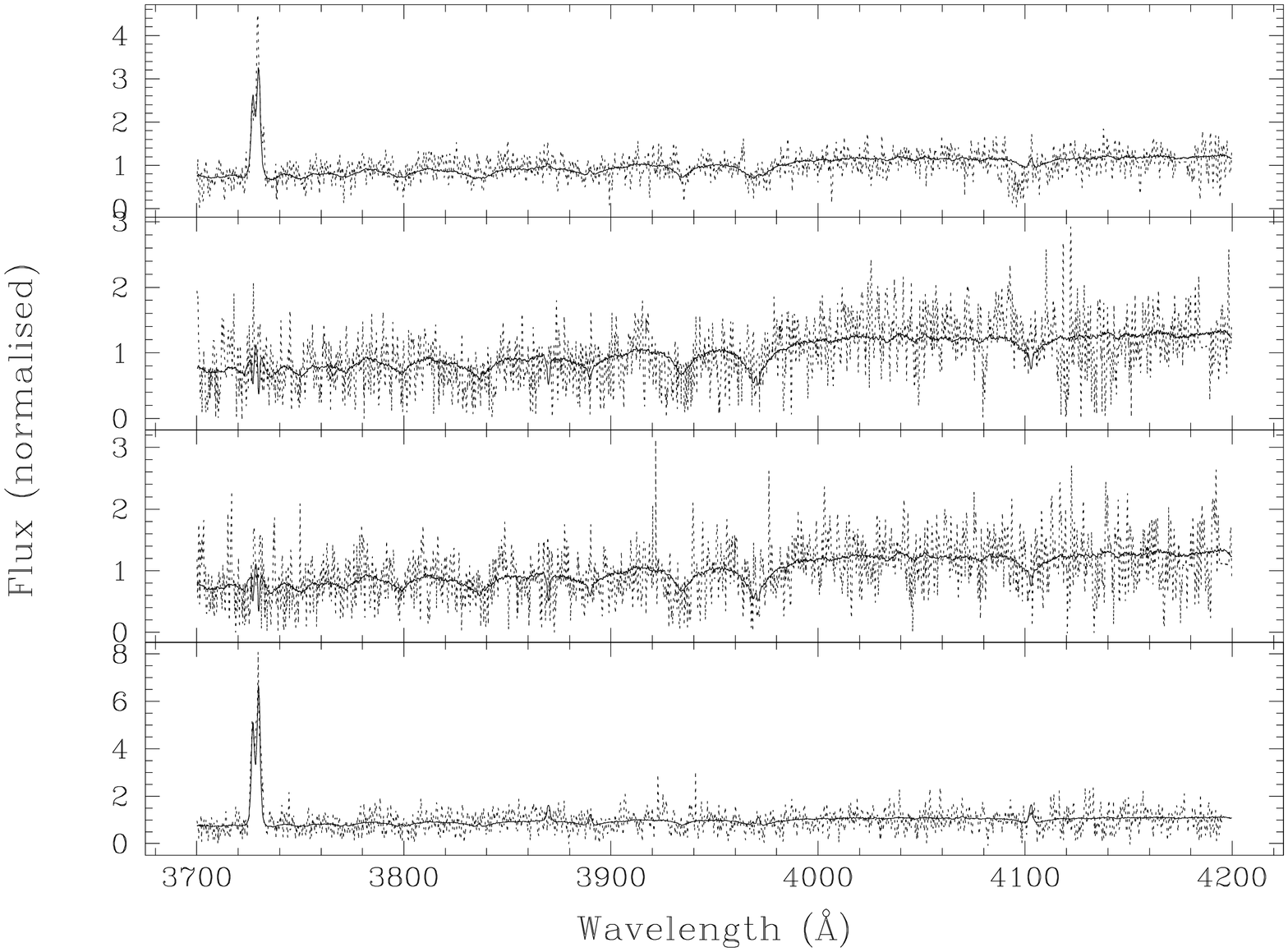,angle=-90,width=3.5in}
\caption{The significance of each principal component is shown in the
top panel, where the improvement in the $\chi^2$ difference between
the galaxy spectrum, 
$f'(\lambda)$, and its reconstruction from the first $i$
components, $\sum_i p_i{\bf{P_i}}$ has been plotted for increasing $i$.  
The lower panel shows how well a set of
randomly chosen galaxy spectra (dotted lines) can be reconstructed
using only the 
mean DEEP2 spectrum and the first principal component, $p_1$ (solid
lines).  This 
highlights the fact that although ${\bf{P}}_1$ appears to only
encapsulate information about the nebular emission features
(Fig.~\ref{fig:evec}), it does
in fact also contain enough information to reconstruct galaxy spectra
without these features.\label{fig:chi} }
\end{center}
\end{figure}

\subsection{Dealing with incomplete data in the PCA}

The treatment of incomplete data in PCA has already been discussed in
the literature, e.g. Everson \& Sirovich (1995) and Connolly \& Szalay
(1999).
However, the
implementation is complicated and deserves further discussion.  The
issues that must be resolved are two-fold:
First, how to determine the eigenspectra (or principal components)
when very few 
galaxies cover the full wavelength range, and second, 
how to project a galaxy onto these eigenspectra when
its spectrum does not cover the entire restframe
wavelength range considered.
It can be shown that the latter issue has a relatively
straightforward, clean solution involving de-correlating the
eigenspectra (which are not orthogonal when we do not use their entire
wavelength coverage).  This has been 
presented in Connolly
\& Szalay (1999).  However, estimating the
eigenspectra in the first place is much more difficult to address.

Consider the situation when the eigenspectra, ${\bf{P}}_{i}$, 
 have
already been estimated using the PCA.  Each of the observed, mean-subtracted
spectra, ${\bf{f'}} = {\bf{f-\bar{f}}}$, 
can then be expressed as a linear combination of these eigenspectra,
\begin{equation}
 {\bf{f'}} = \sum_i p_i {\bf{P}}_{i} \;,
\label{eqn:1}
\end{equation}
where $\bf{\bar{f}}$ is the mean spectrum of all the galaxies and
$p_i={\bf{f'}}\cdot{\bf{P}}_i$ is the projection of each galaxy onto
the $i$th principal component.  The
index $i$ ranges from 1 to the number of resolution elements in each
spectrum.  However, because most of the variance in each
spectrum is contained in only the first few eigenspectra (as will be
demonstrated later), the sum in Eqn.~\ref{eqn:1} can be
truncated to include only the first $k$ most significant elements.

\begin{figure*}
\begin{center}
\epsfig{file=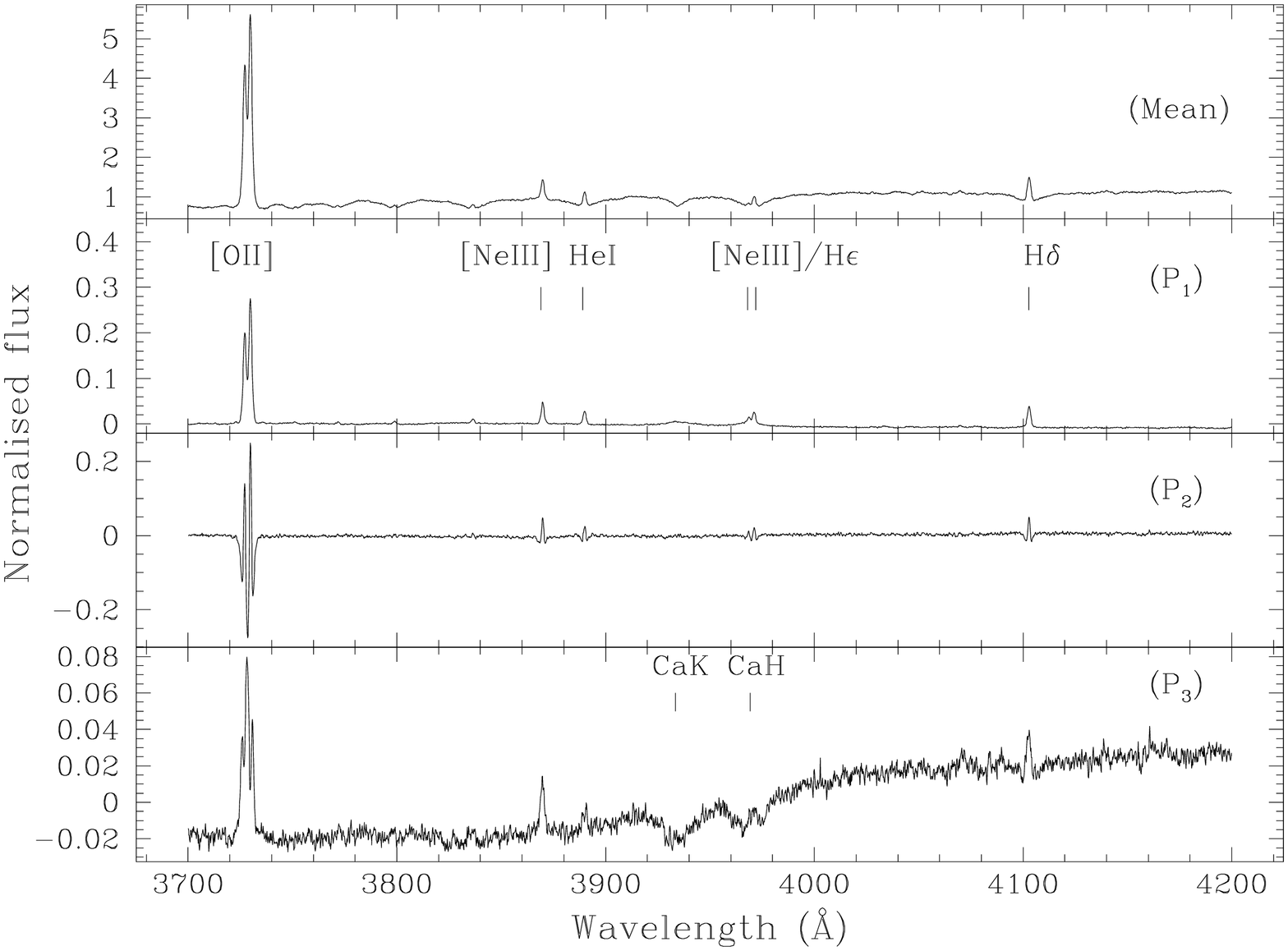,angle=-90,width=3.5in}
\epsfig{file=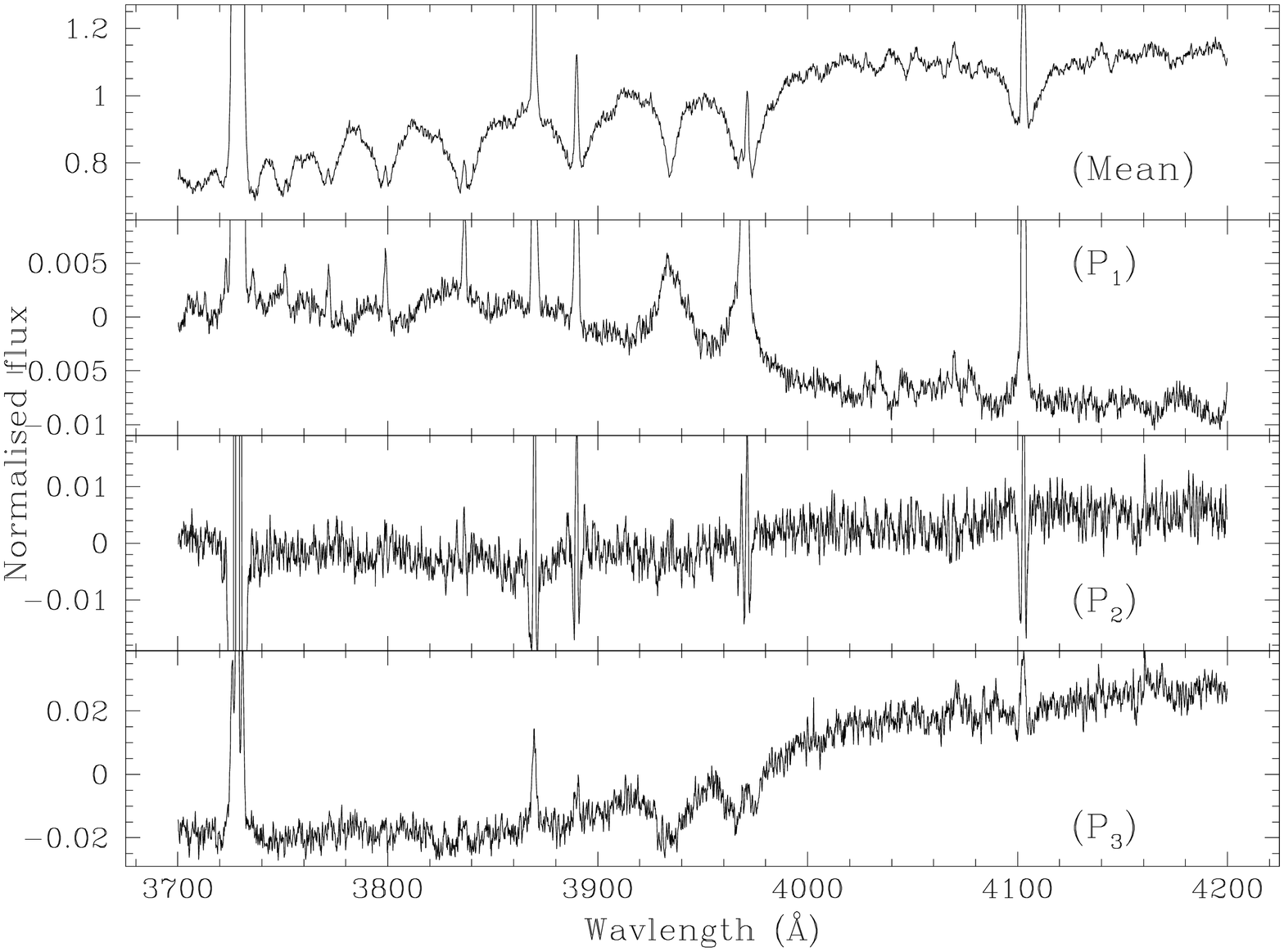,angle=-90,width=3.5in}
\caption{The first three eigenvectors (eigenspectra) derived from the
3700-4200\AA\ PCA are shown (left plot), together with the mean of the
DEEP2 spectra.  The first, ${\bf{P}}_1$, appears to be dominated by the
strength of any emission lines present.  The second displays
artifacts from the varying width of these emission features, while the
third appears to measure the amplitude of the 4000\AA\ break.
Note that these eigenspectra have been
derived from the mean subtracted galaxy spectra, and hence represent
the difference between each galaxy spectrum and the mean spectrum (right
panel).
The lower plot shows a close-up view of the
continuum of the mean spectrum and that of the first 3 eigenspectra,
in which
it can be seen that ${\bf{P}}_1$ also contains an
anti-correlation with the absorption features in each spectrum.  It
can also be seen how each component adds successively
less information about the features present in each spectrum.\label{fig:evec}}
\end{center}
\end{figure*}

We wish to determine how to derive the projections $p_i$ 
when the spectrum, $\bf{f'}$, does not cover the full restframe
wavelength range spanned by the eigenspectra (e.g. because of
poor sky subtraction or the effects of redshift).  The eigenspectra
are defined to be orthonormal,
\begin{equation}
 {\bf{P}}_i \cdot {\bf{P}}_j = \delta_{ij} \;.
\end{equation}
However, when we only consider the wavelength range over which the
observed spectrum, $\bf{f'}$, is defined this is no longer the case.
We denote the incomplete spectrum by $\bf{wf'}$, where $\bf{w}$ is a
vector with non-zero elements only where $\bf{f'}$ is defined.  As
shown in Connolly \& Szalay (1999), the correlations between the
portions of the eigenspectra relevant for $f'$ can be quantified in terms of a
correlation matrix,
\begin{equation}
{\bf{M}}_{ij} = {\bf{wP}}_i \cdot {\bf{P}}_j \;.
\end{equation}
Using this matrix it is possible to show that, when a spectrum does
not cover the entire wavelength range of the eigenspectra,
unbiased estimates of the true projections can be calculated simply
using the inverse of this correlation matrix, $M$,
\begin{equation}
\hat{p}_i = \sum_j {\bf{M}}^{-1}_{ij} {\bf{P}}_j \;,
\label{eqn:deproj}
\end{equation}
where again this sum need only be carried out over the most
significant principal components (in our case we will use the first
$k=20$).  In what follows we use this procedure for two separate purposes:
to estimate the {\em corrected} projections, 
$\hat{p}_i$, when the wavelength range
of the galaxy does not fully span the eigenspectra, and (using these
corrected projections) to interpolate
over any gaps in a spectrum resulting from sky-subtraction etc.,
using Eqn.~\ref{eqn:1}.  This is discussed further in the next section.

\begin{figure}
\plotone{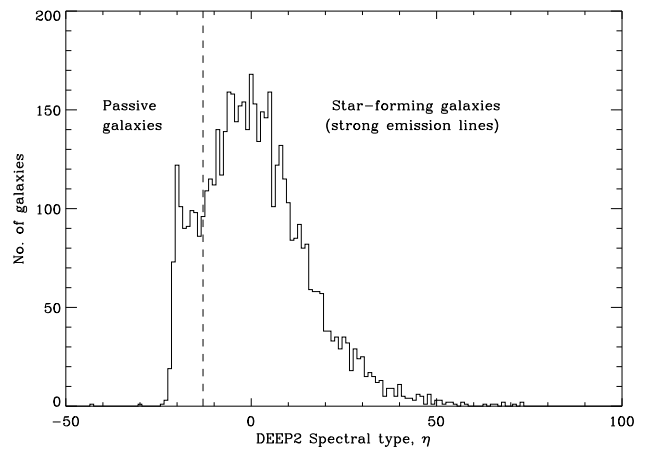}
\caption{The distribution of spectral types, $\eta\equiv p_1$, is shown
for all the galaxies observed to date in the DEEP2 Redshift Survey.
It can be seen that there is a slight bimodality about $\eta =-13$ and
for this reason we adopt a division of our galaxies at this
point.\label{fig:hist}} 
\end{figure}

Before this formalism can be adopted one first requires an estimate of
the eigenvectors of the PCA, ${\bf{P}}_i$.
Several approaches to estimating these eigenspectra in
cases where few or no galaxies cover the full wavelength range of interest
have been suggested. Everson \& Sirovich (1995)  discuss approaches
involving extrapolation of the observed spectra followed by iteration.
Another approach is to make a least-squares
generalization to the E-M PCA (Roweis 1997).  
Each of these methods assumes
that the `gaps' in each spectrum are randomly positioned, which may be
true of low-z galaxy spectra but is clearly unsatisfactory for
high-z surveys since almost all of the
observed spectra will miss a substantial amount of either the extreme 
blue or extreme red end of the full
rest-frame wavelength range. 
For this reason a simple approach to
estimating the eigenspectra will normally fail, since it is difficult to
correlate (and hence extrapolate) one end of the observed spectrum
with what would be expected at the other end.

A more practical solution, which we follow in this paper, is to
restrict the rest-frame wavelength range of interest such that there is
a more significant subsample of galaxies which span 
the entire wavelength range of interest.  
Two such divisions are outlined in Fig.~\ref{fig:aver}; the first involves
considering only the rest-frame wavelength range of
3700-4200\AA\ (covering [O{\sc ii}] and beyond the 4000\AA\  break) 
while the second is extended to 3700-5050 \AA\ (essentially covering
[O{\sc ii}] to [O{\sc iii}]).  There is an obvious
draw-back to this approach in that we are not making use of all the
information contained in our observed spectra; however, as will be
shown below this is a necessary compromise.

\subsection{Limited $\lambda$ PCA}

Using the first wavelength cut (3700-4200\AA), $\sim50$\% of
our galaxy sample span the entire rest-frame wavelength range (with
the exception of small gaps masked out due to bad sky subtraction, gaps
between the CCDs and bad pixels) and
can be used in a PCA analysis.  For the second wavelength cut
(3700-5050\AA), we are using a greater portion of the observed galaxy spectra
and so the analysis is potentially more informative; however, only
$\sim10$\% of galaxies span this entire wavelength range.  

\begin{figure*}
\begin{center}
\epsfig{file=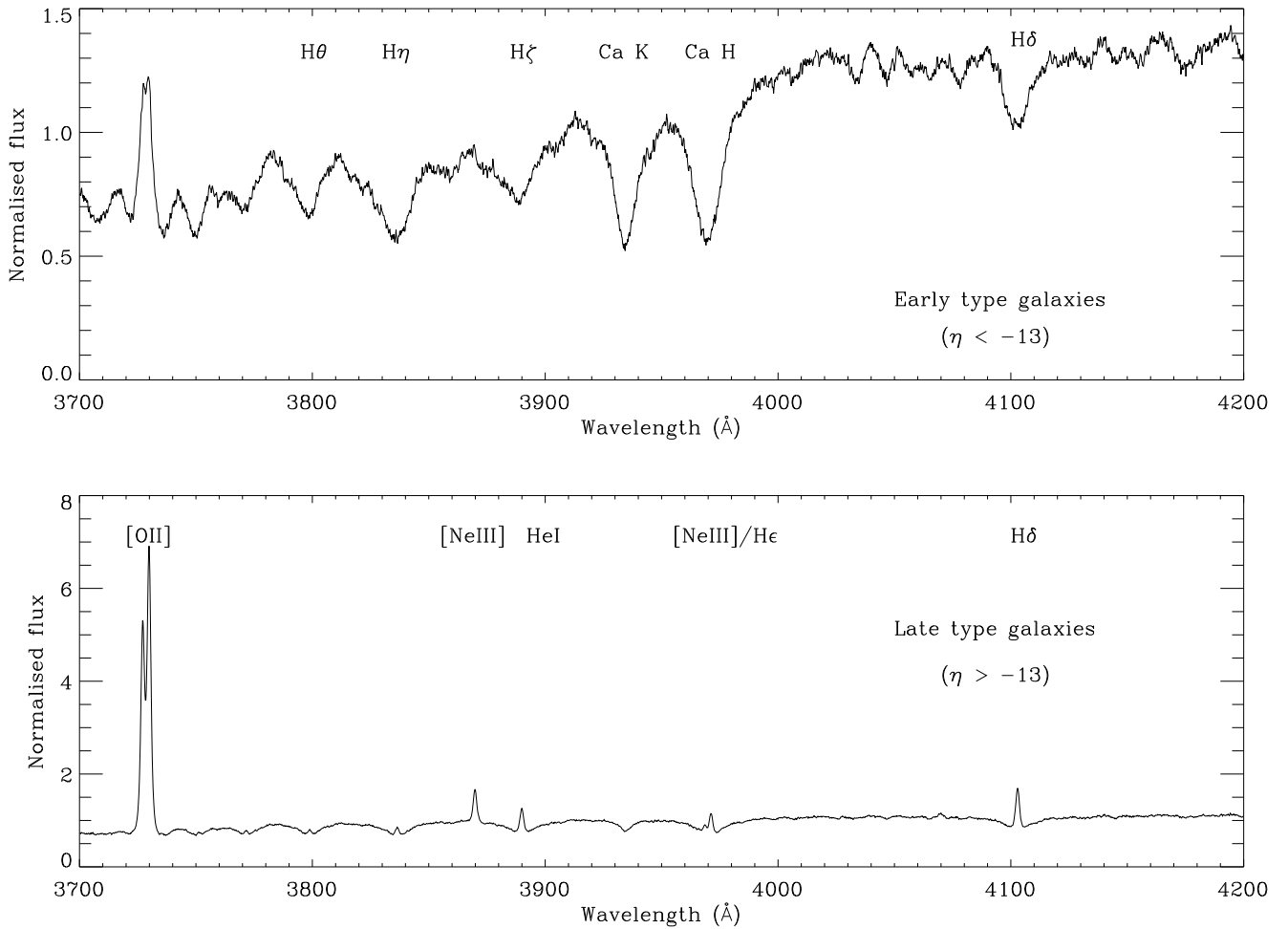,angle=90,width=7in}
\caption{The average spectrum of the two different `types' of galaxies
are shown, again in terms of normalized flux.  It can be seen
that the PCA has been very successful at separating different types of
galaxy spectra.\label{fig:type}}
\end{center}
\end{figure*}

In each case the procedure we adopt is as follows:
\begin{enumerate}
 \item The subset of our galaxies which span the entire restframe
 wavelength range of interest are passed through the E-M PCA to determine
 a first set of $k=20$ eigenspectra, corresponding to the
 orthogonalized columns of the matrix $\bf{C}$ (c.f. Section~3.1).
 \item Gaps in the galaxy spectra used in the previous step are
 interpolated over by re-calculating the `de-correlated'
 projections, $\hat{p}_i$ (as described in the previous section) and 
 are then used to reconstruct the missing parts of the spectrum.
 The new galaxy  spectra are then re-processed through the PCA to
 produce improved estimates of the PCA eigenspectra.
 \item The new set of eigenspectra is then used to compute
 projections for the entire galaxy sample regardless of rest-frame
 wavelength coverage, by means of the de-correlation procedure
 outlined in Eqn.~\ref{eqn:deproj}.
\end{enumerate}
We do not iterate further after projections have been
determined for the entire galaxy sample, since this tends to propagate
errors through the analysis and can lead to unphysical
eigenspectra. 

The values of the first two projections derived in this manner are
plotted in Fig.~\ref{fig:pcs},
for both the rest-frame wavelength ranges considered.  
This figure allows us to determine whether the PCA is robust beyond
the narrow redshift range that corresponds to each of the rest-frame
wavelength limits adopted.  To judge this we have compared in each case the 
distribution of the first two components ($p_1$ and $p_2$) for only
those galaxies used in the initial PCA (the  {\em restricted} $z$ sample), 
and those for the entire
galaxy sample (the {\em unrestricted} $z$ sample).  Although some
small degree of systematic offset is expected, 
the two should be broadly similar, which is clearly not the case for
the second sample used (3700-5050\AA).  This lack of agreement
effectively highlights the limitations of performing this
PCA formalism for spectra that do not have a sufficient degree of overlap.

Clearly the extended sample is more interesting, in so far as it
covers a larger 
rest-frame wavelength range and therefore includes more spectral
features.  However, Fig.~\ref{fig:pcs} demonstrates 
that this 
classification cannot in fact be performed robustly since too few
galaxies cover this full rest-frame wavelength range.  On the other
hand, the more restricted wavelength choice (3700-4200\AA) is much
more robust.  We therefore choose to adopt this
latter wavelength coverage to define our spectral classification.

\section{Spectral Classification}

PCA is not a classification algorithm, rather, it is simply a method of
data compression.  For this reason another (usually manual) step must
be performed, in which the insight gained from this compression is used to
divide a galaxy sample.  This is not necessarily straightforward
and will usually involve some degree of arbitrariness -- especially
since galaxy spectra appear to span a single continuum of types
rather than falling into distinct classes.

By far the most significant output of the PCA is the first principal
component ($p_1$ contains  10 times more variance than any other
component in our analysis, see Fig.~\ref{fig:chi}), which appears to be
dominated by the nebular emission line strengths in a spectrum.  In
particular the strength of the [O{\sc ii}] emission feature figures
prominently, as shown in Fig.~\ref{fig:evec}.  Note however that, as
demonstrated in Fig.~\ref{fig:chi}, this component does in fact also
encapsulate a great deal more information about each spectrum and can be
used to very accurately reconstruct a wide variety of galaxy spectra.
In particular, because the Ca H\&K absorption features are inverted in
this spectrum, it can act as a classifier for both absorption and
emission dominated galaxy spectra.

Other eigenspectra, examples of which are shown in Fig.~\ref{fig:evec},
also contain useful information about each spectrum.  For example, the
second eigenspectrum, ${\bf{P_2}}$, appears to primarily quantify the
width of the various emission lines that is not incorporated into the
first principal component. The component ${\bf{P_3}}$ reflects the stellar
continuum, and in particular the height of the 4000\AA\ break relative
to [O{\sc ii}] that is not already encapsulated in the first principal
component.  
Note that each of these eigenspectra represent variations
in the galaxy spectra relative to the mean spectrum and so, for
example, the [O{\sc ii}] line in ${\bf{P_1}}$ 
represents the additional emission line strength 
present in each spectrum, whether this be less than ($p_1<0$) or
greater than ($p_1>0$) the value for the mean spectrum. 
Beyond the fourth principal component the eigenspectra
become dominated by unphysical features in the galaxy spectra.

Because
the first component is so much more discriminatory between
galaxies (in terms of variance) this appears to be the most logical
(and simplest) classification 
to adopt, particularly since this component alone appears to do well
at reconstructing a large variety of observed spectra.
It is additionally reassuring to note that ${\bf{P_1}}$
also encapsulates the greatest variety of features in each galaxy's
spectrum (Fig.~\ref{fig:evec}), and so is also potentially the most
astrophysically interesting component available from the PCA.
Note that excluding the second component means that we are not making
use of the detailed line-width information, some of which is contained
in $p_2$.  As the DEEP2 reduction pipeline improves more precise
linewidth information, and even detailed rotation curves, will be 
extracted
from the full 2-dimensional spectrum of each galaxy (rather than the
compressed 1-dimensional spectra used here).  In so doing we will be
able to provide a much more accurate characterization of the kinematic
properties of each galaxy that will be complimentary to the
relative emission/absorption line-strengths measured by $p_1$.

\begin{figure}
\plotone{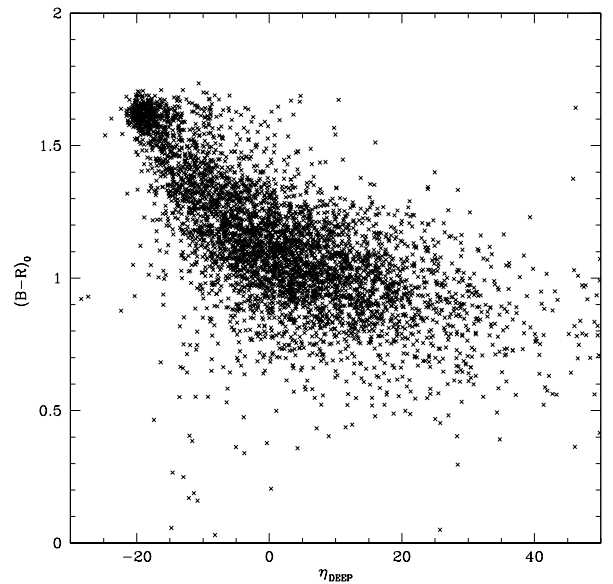}
\caption{The distribution of restframe $(B-R)$ color is compared to
the spectral classification, $\eta_{\rm DEEP}$.  It can be seen that
there is a correlation between the two, particularly for those
galaxies with large $(B-R)$ colors.
\label{fig:color}}
\end{figure}

A similar dominance of the first principal component, $p_1$, has been
noted in previous analyses of low-z
galaxy surveys, suggesting that the distribution of
`normal' galaxy spectra can be well approximated by a one dimensional
sequence. 
However, we must be careful to ensure that this projection is in fact
a robust measure of a galaxy's spectrum if we are to use it alone to
determine spectral type.  For example,
it has been noted for low-z galaxy samples observed through
small fiber optic apertures that the first principal component was not
stable between multiple observations of the same object (Bromley et
al. 1998; Madgwick et al. 2002).  
We find that this is not the case for our most significant
   projections, which proved to have relatively small measurement
   uncertainties.  In particular, we measured the value of $p_1$
   obtained from the instances where a given galaxy spectrum was
   multiply observed and found that the distribution of errors was
   gaussian with a standard deviation of only $\sim$5.
This is most likely owing to the use of
slit-apertures instead of fiber optics.  We therefore choose to adopt
$p_1$ as our measure of spectral type for galaxies observed in the
DEEP2 Redshift Survey and denote it by,
\begin{equation}
 \eta_{\rm DEEP} \equiv p_1 \;.
\end{equation}
This notation has been specifically chosen to analogous to
low redshift samples (e.g. the 2dFGRS) for which a similar
classification scheme has been adopted.\footnote{Note that in the
2dFGRS, $\eta$ was defined as a linear combination of the first {\em
two} components; $p_1$ and
$p_2$.  The two components were needed because of the additional
calibration uncertainties introduced by the small fiber apertures used
in that survey.  Despite this the classifications are in fact comparable
as that combination was also the most statistically significant
output of the PCA which was robust to the known instrumental
uncertainties.}

A histogram of $\eta$ is shown in Fig.~\ref{fig:hist}, in which a significant
bimodality appears to be present.  Since this is the only
discontinuity in the distribution of this projection, it is natural to
divide the sample there ($\eta \sim -13$). 
The presence of this bimodality is interesting in that similar effects
have been observed with galaxy colors (Strateva et al. 2001; Bell et
al. 2003; Weiner
et al. 2003) and also in 
spectral classifications of other samples such as in the 2dFGRS
(Madgwick et al. 2002). 

\begin{figure*}
\epsfig{file=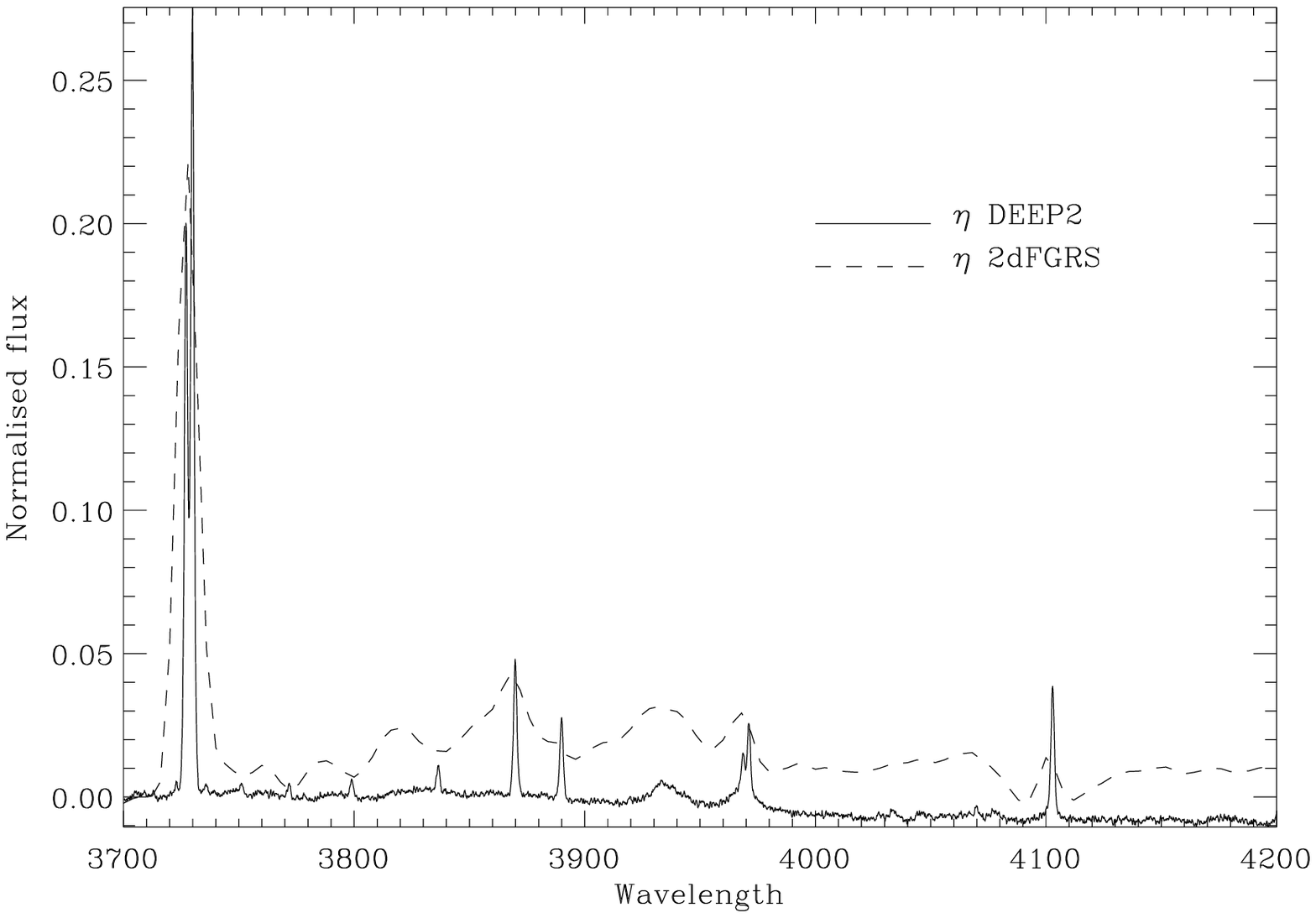,width=3.5in}
\epsfig{file=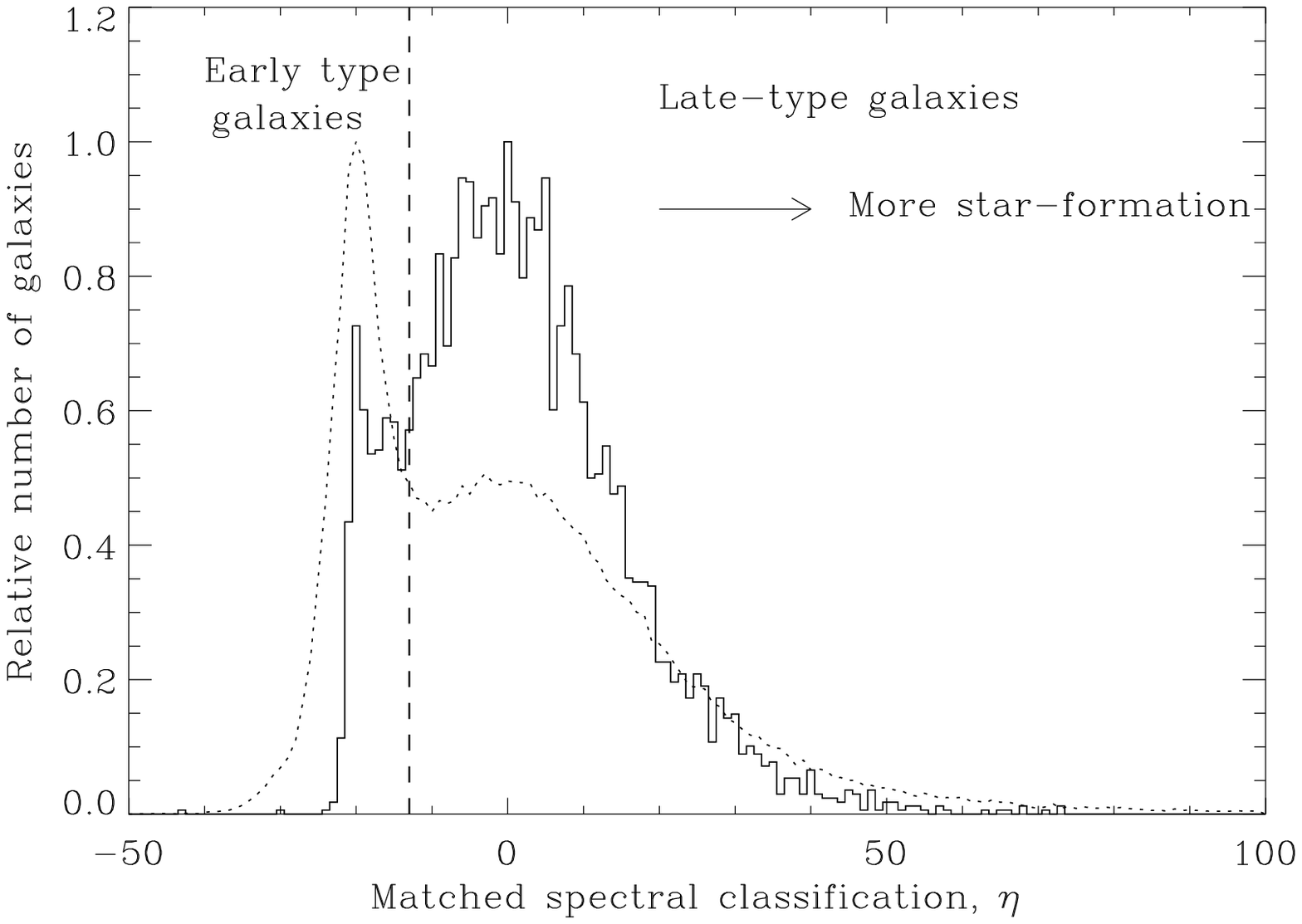,width=3.5in}
\caption{Comparison between the eigenspectra used to classify the
DEEP2 and 2dFGRS spectra (left panel).  It can be seen from this
comparison that the two eigenspectra are using identical features to
classify the galaxies in each survey.  The distributions of the
two spectral types are shown in the right panel (after scaling the
2dFGRS $\eta$ to take account of the different resolutions).
The surveys comprise different types of galaxies based
upon their individual selection criteria, for example it can be seen
that the DEEP2 Survey appears to contain relatively more `late-type'
galaxies.\label{fig:2df}} 
\end{figure*}

Having divided the galaxies into two `types', we proceed to
calculate the mean spectrum for each type, shown in
Fig.~\ref{fig:type}.
Clearly, the difference between these two classes is quite pronounced, and is
differentiating primarily between absorption spectra (associated with older
stellar populations) and emission dominated spectra (associated with
recent star-formation activity).  Taking this interpretation further,
it should be possible to directly relate this classification to the
underlying star-formation activity in each galaxy as was done for the
2dFGRS (see Madgwick et al. 2003).  We return to this point in
Sec.~\ref{sec:2df}. 
 Galaxies with values of
$\eta$ less than $-13$ will be referred to as either `early type'
or passively evolving galaxies, 
whilst those with higher values
will be referred to as `late type' or actively star-forming galaxies.

Figure~\ref{fig:color} compares the 
the restframe $(B-R)_0$ colors of each galaxy with its spectral type,
as defined by our $\eta$ parameter.  
It can be seen from this figure that there is a
correlation between the restframe color and the strength of
emission/absorption features in each observed spectrum, particularly
for those galaxies with the largest $(B-R)_0$.

\section{Comparison with $z=0$ classification}
\label{sec:2df}

In this Section we briefly compare the classification derived for the DEEP2
Survey with that
adopted in the 2dF Galaxy Redshift Survey (Colless et al. 2001), which
comprises over 200,000 low redshift galaxy spectra.
The reason we make this particular comparison is that
the two classification regimes have been similarly defined, in that
they are both derived from the most significant component of the PCA
that is robust to the known instrumental uncertainties (Madgwick et
al. 2002).   Such a
classification for the Sloan Digital Sky Survey is also in progress
but is not yet available.

Both classification methods provide a continuous parameterization of
the spectral type of a galaxy, denoted by $\eta$.  At a practical
level, this spectral
classification is simply a dot-product between each observed galaxy
spectrum 
and the chosen classifying eigenspectra, which are shown
in Fig.~\ref{fig:2df} for the two surveys.  
It can be seen from this figure that despite
the very different selection criteria in the two surveys, the chosen
classifications both sample the same spectral features in the same
relative proportions over their common wavelength interval.  Hence the
two classifications should be very comparable for broad studies of
galaxy evolution.
 
The distributions of the $\eta$ spectral types are also shown in
Fig.~\ref{fig:2df}.  
Note however that because the DEEP2
survey spectra have much higher resolution, the locus of projections
for this survey spans a much larger range (since each spectrum has
been normalized to have mean flux of unity regardless of resolution).  
For this reason we have scaled
the $\eta_{\rm 2dF}$ spectral parameter of the 2dFGRS survey to match
$\eta_{\rm DEEP}$ in this comparison.  We find that multiplying by
a factor of 8 suffices to match the two classifications,
a similar factor would be derived by noting that early type galaxies
are separated from late types at
$\eta_{\rm 2dF}=-1.4$ (Madgwick et al. 2002), as opposed to $\eta_{\rm
DEEP}=-13$.

\begin{figure}
\epsfig{file=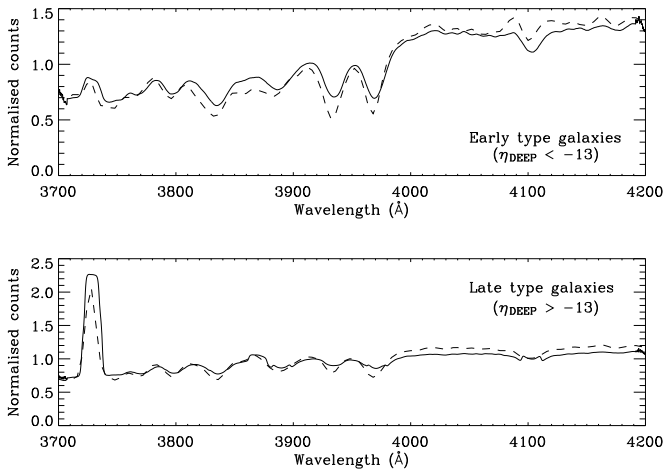,width=3.5in}
\caption{The average spectrum of the
early and late type galaxies from the 2dFGRS and DEEP2 surveys are
compared.  The DEEP2 galaxy spectra (solid line) have been 
smoothed using an 8\AA\ filter in order to match the resolution of the
2dFGRS spectra. \label{fig:2df_spec}}
\end{figure}

Figure~\ref{fig:2df_spec} shows a comparison of the average
spectra for early and late type galaxies as calculated from the 2dFGRS
and DEEP2 spectra (with the latter smoothed to the 9\AA\ resolution of
the former).  This figure again highlights that the 
correspondence between the two classifications is quite
striking, despite being derived over different wavelength
ranges and at different resolutions, they essentially encapsulate the
same physical information.

It has already been shown (Madgwick et al. 2003)
that the spectral classification $\eta$ adopted for the 2dFGRS
corresponds most naturally to the relative amount of star formation
activity currently occurring in each galaxy as compared with its past
average (the Scalo birthrate, $b$, parameter, Scalo 1986). Objects with high
$\eta$ values are galaxies with particularly
strong recent star formation, whereas the lowest-$\eta$ sample of galaxies 
has
$b\le0.1$ (i.e. their current star formation rate
is only 10\% of their past averaged value).  
Because there is
such a simple one-to-one correspondence between this spectral
classification and the star formation activity of each galaxy, we can
confirm from Fig.~\ref{fig:2df} our intuition that selecting galaxies
using restframe $U$ magnitudes (as is the case for the DEEP2 Survey,
see e.g. Weiner et al.~2003) is very much more biased towards
galaxies with recent episodes of star-formation than the $B$ selection
adopted for the 2dFGRS.
This result will have important repercussions for the interpretation
of future analyses of the DEEP2 Survey data.

A more
detailed assessment of the exact relative frequency of star forming
galaxies in each survey will be forthcoming in a future paper, in
which the selection function of each survey will be fully
incorporated.  Once this is available it will also be particularly
interesting to 
experiment with how consistent the evolution between the populations
is with that expected from spectral synthesis models (e.g. Bruzual \&
Charlot 1993; Fioc \& Rocca-Volmerange 1997).

\section{Conclusions}

In this paper we have presented a new PCA-based spectral
classification, $\eta_{\rm DEEP}$, 
for the galaxies observed to date in the DEEP2 Redshift 
Survey.  The main goal in developing this classification was to
provide a consistent and robust measure of the type of a galaxy
over the large range of redshifts encountered in this survey.
To do this special handling of incomplete and `gappy' observed spectra
is required.

Although this classification, $\eta_{\rm DEEP}$, appears to primarily
identify galaxies with differing strengths of nebular emission,
Fig.~\ref{fig:chi} demonstrates that this component alone can
reconstruct the spectra of a wide variety of galaxies extremely well.  

We have not directly related this
classification to the role of star formation activity in a given galaxy
However, there is strong evidence from a similar analysis
carried out for the 2dFGRS that $\eta_{\rm DEEP}$ should also
correlate well with
the relative amount of star formation.
A more detailed study of this correlation, together with the
role of higher order PCA projections, will be forthcoming when higher
resolution spectral synthesis models become publicly available in the
near future (the DEEP2 galaxy spectra are at a resolution that is $\sim20$
times higher than any synthesis model currently available).

This particular classification will be particularly
useful in subsequent analyses e.g. of the galaxy luminosity function
or correlation functions, as it is easily comparable with other
classifications at $z=0$, for which large spectroscopic samples are
now publicly available.  In addition, previous work (e.g. Madgwick
et al. 2003) has shown that it is straightforward to make direct
comparisons between such a spectral classification regime and the
output of semi-analytic galaxy models (e.g. Kauffmann, White \&
Guiderdoni 1993; Cole et al. 1994; Somerville \& Primack 1999),
allowing us to directly constrain the assumed models of 
galaxy formation and evolution between $z=1$ and $z=0$ using this form
of classification.  

When used in conjunction with spectral synthesis models, we expect 
that parameters similar to the one presented here
will provide a wealth of information on both the amount
and `type' of evolution that has occurred in the galaxy population.  This
is particularly relevant now that such large 
samples of galaxy spectra are becoming available over such 
a wide range of
redshifts.  For this reason we expect studies of the evolution in
the galaxy population to become a particularly rich field of research
in the near future, and that the DEEP2 Redshift Survey will play an
especially prominent role in this field.

\acknowledgements

This work was supported in part by NSF grants AST00-71048
and KDI-9872979.   The DEIMOS spectrograph was funded by a grant from CARA
(Keck Observatory), by an NSF Facilities and Infrastructure grant
(AST92-2540),  
by the Center for Particle Astrophysics,
and by gifts from Sun Microsystems and the Quantum Corporation.
DPF is supported by a Hubble Fellowship.

The DEEP2 Redshift Survey has been made possible through the dedicated
efforts of the DEIMOS staff at UC Santa Cruz who built the instrument
and the Keck Observatory staff who have supported it on the telescope.

\end{document}